\documentclass[twocolumn]{aastex701}

\usepackage{amsmath}
\usepackage{subfigure}

\begin{document}

\title{A Hierarchical Bayesian Differential Calibration of the
\textit{Gaia} DR3 Parallax Zero Point Using Galactic Globular Clusters}

\author{Ye Ding}
\affiliation{Shanghai Astronomical Observatory, Chinese Academy of Sciences, Shanghai 200030, China}
\affiliation{University of Chinese Academy of Sciences, Beijing 100049, China}
\email{yeding@shao.ac.cn}

\author{Shilong Liao} 
\affiliation{Shanghai Astronomical Observatory, Chinese Academy of Sciences, Shanghai 200030, China}
\affiliation{University of Chinese Academy of Sciences, Beijing 100049, China}
\email{shilongliao@shao.ac.cn}

\author{Qi Xu} 
\affiliation{Shanghai Astronomical Observatory, Chinese Academy of Sciences, Shanghai 200030, China}
\affiliation{University of Chinese Academy of Sciences, Beijing 100049, China}
\email{qixu@shao.ac.cn}

\author{Qiqi Wu} 
\affiliation{Shanghai Astronomical Observatory, Chinese Academy of Sciences, Shanghai 200030, China}
\affiliation{University of Chinese Academy of Sciences, Beijing 100049, China}
\email{wuqiqi@shao.ac.cn}

\author{Keyu Zhu} 
\affiliation{Shanghai Astronomical Observatory, Chinese Academy of Sciences, Shanghai 200030, China}
\affiliation{University of Chinese Academy of Sciences, Beijing 100049, China}
\email{zhukeyu@shao.ac.cn}

\author{Zhaoxiang Qi}
\affiliation{Shanghai Astronomical Observatory, Chinese Academy of Sciences, Shanghai 200030, China}
\affiliation{University of Chinese Academy of Sciences, Beijing 100049, China}
\email{zxqi@shao.ac.cn}

\correspondingauthor{Shilong Liao}
\email{shilongliao@shao.ac.cn}


\begin{abstract}

Globular clusters (GCs) occupy a broad range of stellar colors and magnitudes while sharing nearly the same distance. \textit{Gaia} Data Release 3 parallaxes retain systematic zero-point residual variations for GCs, even after the correction by the widely used standard correction model (L21). We derive an empirical residual correction by the Hierarchical Bayesian Model, separating cluster-level offsets from the differential information carried by the member sequences. We first evaluate the L21 baseline at a fixed reference color, which assigns the remaining color-dependent correction to the residual model. Then we model the residual zero point, $\delta Z(G,\nu_{\rm eff})$, with B-splines functions in color and magnitude--color interaction terms. The uncorrected cluster residuals exhibit a negative offset of approximately $25$--$40~\mu{\rm as}$. The L21 baseline removes most of this offset but leaves coherent color- and magnitude-dependent structure. Applying the residual model brings the mean residuals close to zero, especially across most of the sampled color range. In five-fold cross-validation that excludes entire clusters from training, the correction modestly reduces the magnitude- and color-trend RMS relative to the L21 correction. The resulting model therefore provides a differential refinement of the \textit{Gaia} DR3 parallax zero-point within the calibrated domain for GCs, rather than an independent determination of its absolute value.
\end{abstract}



\section{Introduction}
The European Space Agency's \textit{Gaia} mission has delivered astrometry of unprecedented precision for more than a billion sources in the Milky Way and beyond \citep{2016A&A...595A...1G,2021A&A...649A...1G}. Nevertheless, the \textit{Gaia} Data Release 3 (DR3) parallaxes retain small but astrophysically important systematic errors. The most prominent is the parallax zero-point offset, which arises from effects including basic-angle variations, charge-transfer inefficiency, and changes in the CCD gating and windowing configurations \citep{2021A&A...649A...2L}. Its amplitude ranges from a few to several tens of $\mu$as and becomes important when many sources are combined or when high-precision distances are required.

\citeauthor{2021A&A...649A...4L} (\citeyear{2021A&A...649A...4L}, hereafter L21) constructed the standard correction model as a global function of $G$ magnitude, effective wavenumber $\nu_{\rm eff}$, and ecliptic latitude $\beta$, using quasars and other reference samples. Although this correction removes much of the global parallax bias, residual structure remains ( \citealt{2026MNRAS.550g1115D}, hereafter D26). Its accuracy is limited by the coverage of the calibration samples in the color--magnitude plane and by spatial variations that cannot be represented fully by a smooth global function. These limitations are particularly relevant for sources at the boundaries of the calibrated color range and for crowded stellar fields.

Globular clusters (GCs) provide a complementary way to examine these limitations. Their members occupy a broad range of stellar colors and magnitudes while sharing nearly the same distance. Because the members of a given cluster are also confined to a relatively small area of the sky, large-scale position-dependent errors are approximately common to them. Trends in their relative parallaxes can therefore reveal source-dependent residuals without requiring an independently known cluster distance. This makes GCs particularly useful for testing whether a global correction transfers to the stellar populations and crowded fields that are less directly represented by its reference samples.

\begin{figure}
    \centering
    \subfigure{\includegraphics[width=1.0\linewidth]{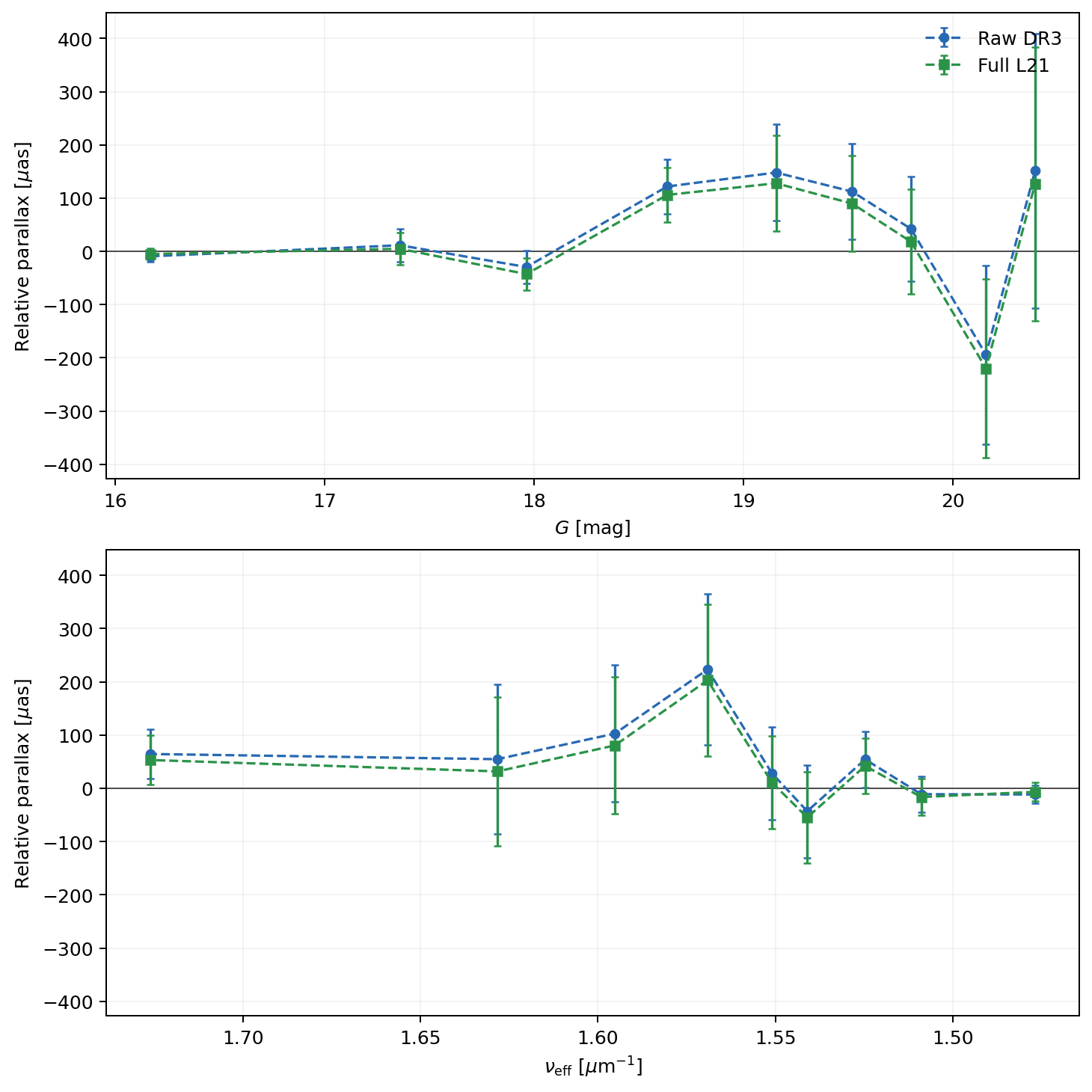}}
    \caption{Parallax residual trends for members of NGC~7492 (N=187) for the raw (blue) and corrected residuals by L21 (green). The top and bottom rows show residuals as functions of $G$ magnitude and effective wavenumber $\nu_{\rm eff}$, respectively. The residuals are measured relative to the mean parallax shown in the inset. }
    \label{fig:NGC_7492_residual_trends}
\end{figure}

However, D26 found the significant residual trends with $G$ and $\nu_{\rm eff}$ among the members of GCs, after applying several global corrections including L21 and D26. Figure~\ref{fig:NGC_7492_residual_trends} illustrates this behavior for NGC~7492. These behavior highlight a critical limitation inherent to any generalized, all-sky parallax zero point calibration model. The global corrections trained on a vast, diverse population of field stars (QSOs, binaries, etc.) are not applicable to GCs which are highly homogeneous populations with distinct and tightly constrained distributions in the color–magnitude domain.

In this work, rather than treating all GC member stars as a homogeneous data set, we formulate a Hierarchical Bayesian Model \citep{2014bda..book.....G,2019PASA...36...10T}. Each cluster has its own effective parallax and additional scatter, whereas all clusters share a residual zero-point function of color and magnitude. This structure separates cluster-level offsets from the differential information carried by the member sequences. Our aim is not to replace a full-sky \textit{Gaia} zero-point model or to determine an independent absolute parallax zero-point. Instead, we test whether cluster members reveal coherent color--magnitude trends that remain after a global correction and quantify those trends over the domain sampled by the calibration stars. 

The paper is organized as follows. Section~\ref{sec:data} describes the cluster samples. Section~\ref{sec:model} presents the hierarchical model and the B-spline representation of the residual zero-point. Section~\ref{sec:result} reports the fitted correction and the internal residual diagnostics. Section~\ref{sec:validation} presents the held-out GC validation, and Section~\ref{sec:discussion} examines sensitivity to the model specification, investigate the application for open clusters, and discuss the model limitations. Section~\ref{sec:summary} summarizes our conclusions.

\section{Data}\label{sec:data}

\subsection{Stellar clusters as differential calibrators}

High-probability members of a sufficiently compact stellar cluster share nearly the same geometric parallax. Residual differences among member parallaxes as functions of magnitude or color can therefore constrain parameter-dependent zero-point errors. This approximation becomes less accurate for nearby or spatially extended clusters, whose line-of-sight depth can produce a measurable intrinsic parallax dispersion. Following \citet{2021A&A...649A...5F}, we require the catalog or literature distance estimate to satisfy $d_{\rm ref}>1~{\rm kpc}$. 
The common-parallax assumption constrains only differential variations within a cluster. A constant offset shared by all members is degenerate with the cluster parallax and is absorbed into the cluster-specific intercept introduced in Section~\ref{sec:model}. Consequently, the cluster data do not independently determine an additional absolute zero-point constant.

\begin{figure*}
    \centering
    \subfigure{\includegraphics[width=0.9\linewidth]{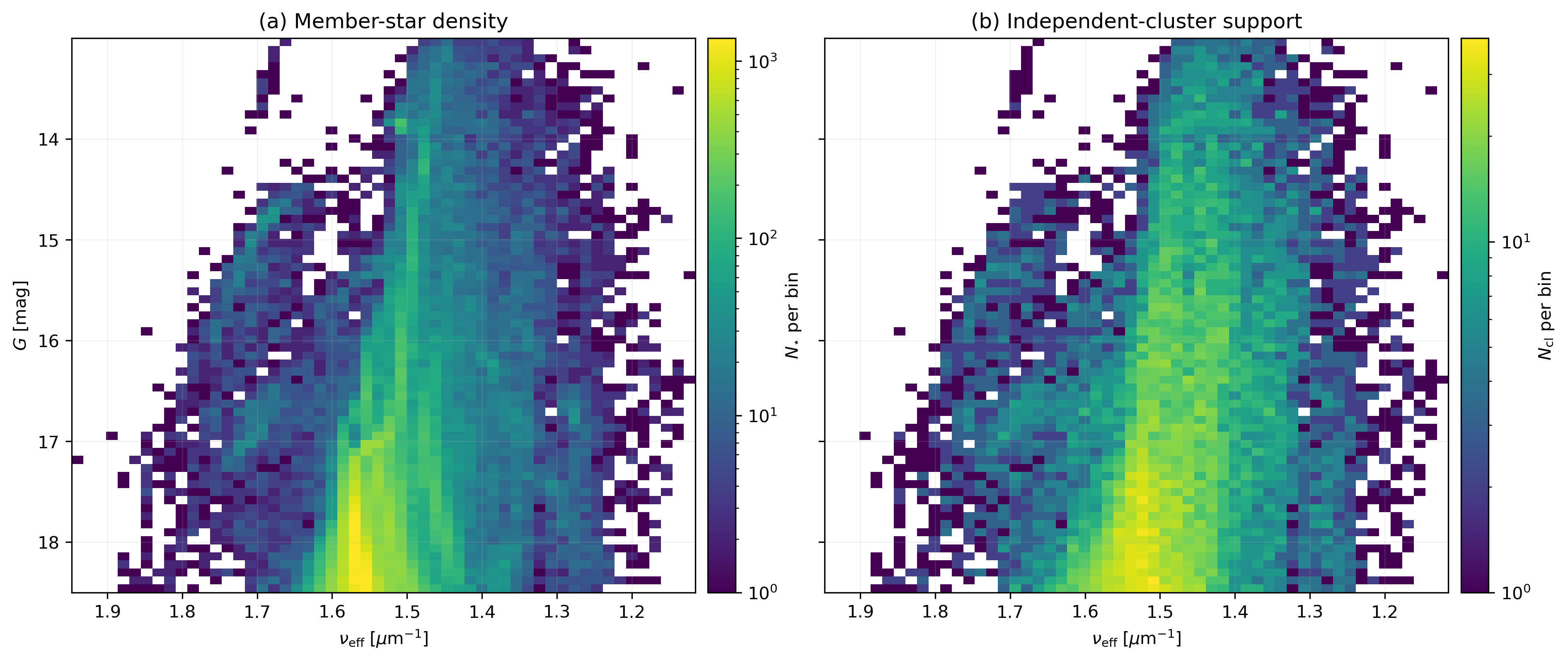}}
    \caption{Color--magnitude coverage of the final cluster-member sample. Both panels show the distribution of retained five-parameter Gaia sources in the $(G,\nu_{\rm eff})$ plane. Panel (a) shows the number of member stars per bin, $N_\star$, while panel (b) shows the number of independent cluster systems contributing to each bin, $N_{\rm cl}$. The color scales are logarithmic, and empty bins are shown in white. The second panel distinguishes regions dominated by a few rich clusters from regions supported by many independent cluster systems.}
    \label{fig:cmd_clusters}
\end{figure*}

\begin{figure}
    \centering
    \subfigure{\includegraphics[width=1.0\linewidth]{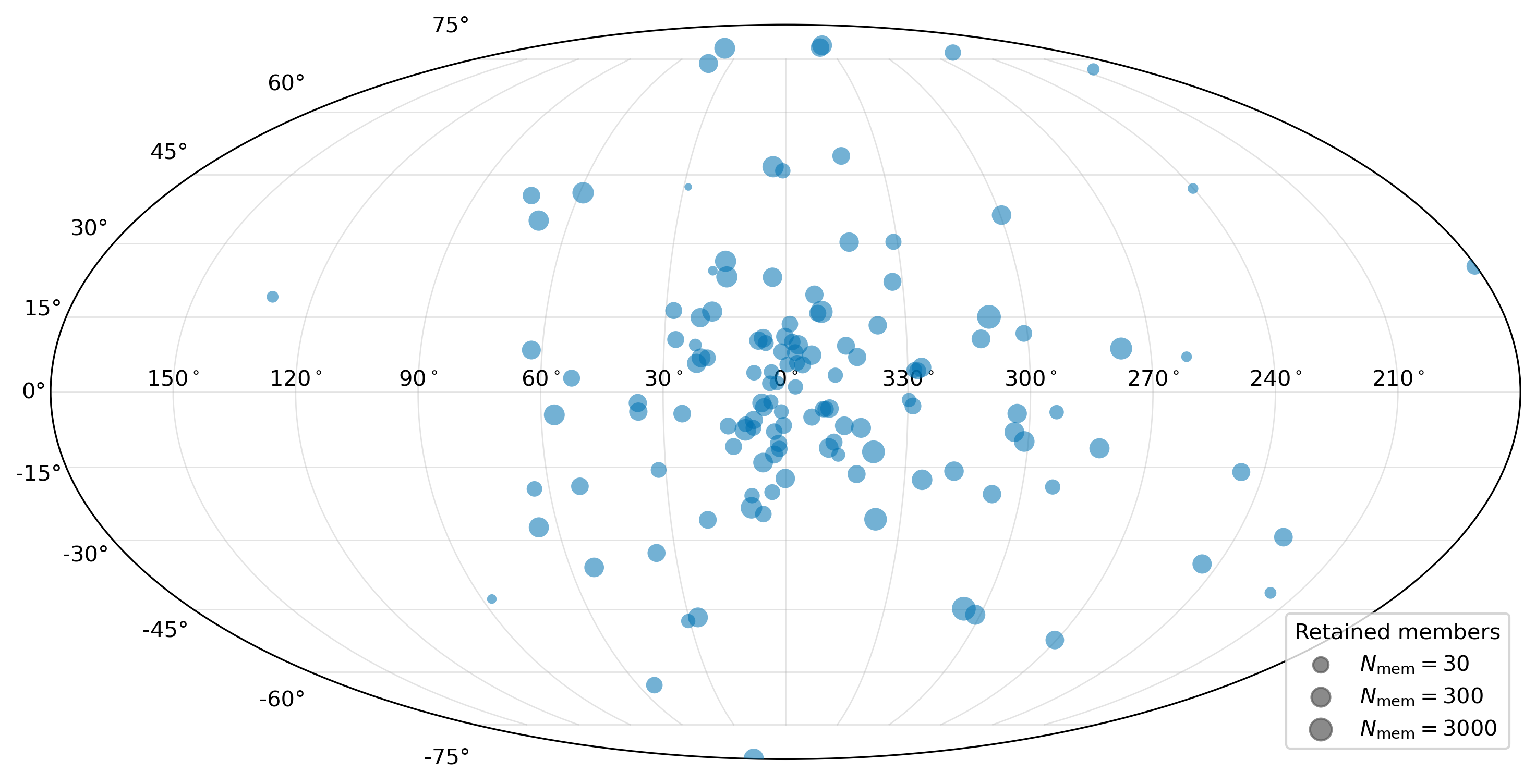}}
    \caption{All-sky distribution of the retained cluster systems in Galactic coordinates, shown in a Mollweide projection. Each point marks the center of one cluster. Symbol area increases with the number of retained member stars, $N_{\rm mem}$.} 
    \label{fig:sky_position}
\end{figure}

\subsection{Sample selection}
For the GC sample, we adopt the member catalog of \citet{2021MNRAS.505.5978V}, which provides membership classifications for 170 Galactic clusters. We retain only sources with five-parameter astrometric solutions so that the calibration uses the photometric effective wavenumber $\nu_{\rm eff}$ consistently. Six-parameter solutions are excluded because their calibration depends on a pseudo-color and follows a distinct zero-point relation. We require the member probability, $P_{\rm mem}>0.9$, and $qflag=3$, corresponding to the highest-quality astrometric selection defined by \citet{2021MNRAS.505.5978V}. What's more, as the overestimated mean BP flux for faint sources in the published photometry \citep{2021A&A...649A...3R}, we only retain the members with $G <$18.5. We also require at least 20 retained members per cluster to ensure the stability of the model fit. The final calibration sample contains 107,222 stars in 138 clusters.

Figure~\ref{fig:cmd_clusters} shows the color--magnitude coverage of the final cluster-member sample in the $(G,\nu_{\rm eff})$ plane. The sample spans approximately $13 \lesssim G \lesssim 18.5$ mag and $1.1 \lesssim \nu_{\rm eff} \lesssim 1.9~\mu{\rm m}^{-1}$, with the highest density occurring over the main color--magnitude region populated by the selected cluster members. The broad coverage in both magnitude and effective wavenumber allows the residual zero-point function to be constrained over the parameter domain most relevant to the training sample. The left panel of Fig.~\ref{fig:cmd_clusters} shows the number of member stars in each bin and therefore indicates the statistical weight contributed to the likelihood, while the right panel shows the number of independent cluster systems contributing to each bin. The regions of high member-star density are generally also supported by many clusters, whereas the brightest and the most extreme color regions are sampled by relatively few systems. Figure~\ref{fig:sky_position} shows the distribution of the retained cluster centers in Galactic coordinates.

\section{Bayesian Hierarchical Framework}
\label{sec:model}

\begin{deluxetable*}{lll}
\tablewidth{0pt}
\tabletypesize{\scriptsize}
\tablecaption{Symbols used in the hierarchical residual parallax
zero-point model.\label{tab:symbols}}
\tablehead{
\colhead{Symbol} & \colhead{Unit} & \colhead{Definition}
}
\startdata
$\mathcal{L}$ & --- & Likelihood function \\
$\pi$ & --- & Prior density \\
$p(\boldsymbol{\phi}\mid\mathcal{D})$ & --- & Posterior density \\
$\boldsymbol{\phi}$ & --- & Full parameter vector \\
$\mathcal{D}$ & --- & Full set of observed member-star parallaxes and covariates \\
$i$ & --- & Index of a member star \\
$k$ & --- & Index of a stellar cluster \\
$K$ & --- & Number of clusters in the final sample \\
$N_k$ & --- & Number of retained members in cluster $k$ \\
$\hat{\varpi}_{i}^{(k)}$ & mas & Published \textit{Gaia} parallax of star $i$ in cluster $k$ \\
$\sigma_{\varpi,i}$ & mas & Formal \textit{Gaia} parallax uncertainty of star $i$ \\
$\varpi_{{\rm true},k}$ & mas & True geometric parallax of cluster $k$ \\
$\eta_k$ & mas & Common local parallax offset across cluster $k$ \\
$\varpi_{c,k}$ & mas & Effective cluster parallax,
$\varpi_{c,k}=\varpi_{{\rm true},k}+\eta_k$ \\
$Z_{0,i}$ & mas & Color-fixed L21 baseline correction for star $i$ \\
$\delta Z(G,\nu_{\rm eff})$ & mas & Residual parameter-dependent parallax zero-point correction \\
$G_i$ & mag & \textit{Gaia} $G$-band magnitude of star $i$ \\
$\nu_{{\rm eff},i}$ & $\mu{\rm m}^{-1}$ & Effective wavenumber of star $i$ \\
$\theta_j$ & mas & Coefficient of the $j$th residual B-spline basis function \\
$\epsilon_i^{(k)}$ & mas & Star-specific random residual term \\
$\nu_{\rm ref}$ & $\mu{\rm m}^{-1}$ & Reference effective wavenumber used to evaluate $Z_0$ \\
$\beta_i$ & deg & Ecliptic latitude of star $i$ \\
$r_i^{(k)}$ & mas & Baseline-corrected residual,
$r_i^{(k)}=\hat{\varpi}_i^{(k)}-Z_{0,i}$ \\
$s_k$ & mas & Cluster-specific extra-scatter scale \\
$V_{i,k}$ & mas$^2$ & Total variance, $V_{i,k}=\sigma_{\varpi,i}^2+s_k^2$ \\
$M_p(G)$ & --- & Cubic B-spline basis function in magnitude \\
$C_q(\nu_{\rm eff})$ & --- & Cubic B-spline basis function in effective wavenumber \\
$\Psi_j(G,\nu_{\rm eff})$ & --- & Retained color or magnitude--color basis function \\
$P$ & --- & Number of retained magnitude contrasts \\
$Q$ & --- & Number of retained color contrasts \\
$M$ & --- & Total number of residual coefficients, $M=Q+PQ$ \\
$\tau_{\varpi}$ & mas & Scale of the prior on the effective cluster parallax, $\varpi_{c,k}$ \\
$\sigma_s$ & mas & Scale of the Half-Normal prior on $s_k$ \\
$\sigma_\theta$ & mas & Standard deviation of the Gaussian prior on $\theta_j$ \\
\enddata
\end{deluxetable*}

The calibration problem is naturally expressed as a Hierarchical Bayesian Model \citep{2017ARA&A..55..213S,2019PASA...36...10T}. At the stellar level, each measured parallax is described by its formal \textit{Gaia} uncertainty and a cluster-specific additional scatter. At the cluster level, all retained members share a common effective parallax. At the population level, a smooth polynomial function describes the residual zero point shared by all clusters. This hierarchy separates the within-cluster differential information that constrains the shape of the correction from cluster-level offsets that are not identifiable as zero-point errors.

Bayesian inference combines the likelihood of the member-star parallaxes with weakly informative priors on the cluster parameters and regularization of the flexible spline correction \citep{2014bda..book.....G}. Denoting all model parameters by $\boldsymbol{\phi}$ and the full data set by $\mathcal{D}$, the posterior is
\begin{equation}
p(\boldsymbol{\phi}\mid\mathcal{D}) \propto
\mathcal{L}(\mathcal{D}\mid\boldsymbol{\phi})\,
\pi(\boldsymbol{\phi}),
\label{eq:posterior_general}
\end{equation}
where the likelihood $\mathcal{L}$ describes the \textit{Gaia} measurements and the prior $\pi$ supplies weak regularization for model parameters. The complete parameter vector is $\boldsymbol{\phi}=\{\varpi_{{\rm c},k},s_k,\theta\}$, where $\varpi_{{\rm c},k}$ and $s_k$ are the effective parallax and additional scatter of cluster $k$, respectively, and $\theta$ contains the residual zero-point coefficients.

\subsection{Observation model}
Let $\hat{\varpi}_i^{(k)}$ denote the observed \textit{Gaia} DR3 parallax of member star $i$ in cluster $k$, and let $\sigma_{\varpi,i}^{(k)}$ denote its reported formal uncertainty. For
a sufficiently distant and compact cluster, its high-probability members are expected to share an approximately common geometric parallax. However, all members may also be affected by a local
astrometric offset that cannot be distinguished from the cluster parallax using the member data alone. We therefore define the effective cluster intercept as
\begin{equation}
    \varpi_{c,k}
    =
    \varpi_{{\rm true},k}+\eta_k,
    \label{eq:effective_cluster_parallax}
\end{equation}
where $\varpi_{{\rm true},k}$ is the true geometric parallax of cluster $k$ and $\eta_k$ represents an unresolved cluster-level astrometric offset. These two components are not separately identifiable from the member-star parallaxes; throughout this work, we infer only their sum,
$\varpi_{c,k}$.

Under our sign convention, the observed parallax is written as
\begin{equation}
    \hat{\varpi}_i^{(k)}
    =
    \varpi_{c,k}
    + Z_{0,i}
    + \delta Z
      \left(G_i,\nu_{{\rm eff},i};\boldsymbol{\theta}\right)
    + \epsilon_i^{(k)},
    \label{eq:raw_observation_model}
\end{equation}
where \(Z_{0,i}\) is the adopted global baseline, $\delta Z$ is the residual color--magnitude-dependent zero-point term, and $\epsilon_i^{(k)}$ represents the remaining star-level measurement
error. 

We define the baseline using the L21 correction for five-parameter
\textit{Gaia} solutions,
\begin{equation}
    Z_{0,i}
    =
    Z_5(G_i,\nu_{\rm ref},\beta_i),
    \label{eq:fixed_colour_baseline}
\end{equation}
evaluated at $\nu_{\rm ref}=1.59~\mu{\rm m}^{-1}$. This value is approximately the median effective wavenumber of the quasar sample used in the L21 calibration and therefore places the reference color near the center of the original calibration sample. The resulting $Z_0$ retains the
large-scale dependence of L21 on magnitude and ecliptic latitude, while the source-dependent color structure is assigned explicitly to $\delta Z(G,\nu_{\rm eff})$. Changing $\nu_{\rm ref}$ shifts the fixed-color L21 baseline and consequently changes the fitted residual coefficients. For a sufficiently flexible residual model, however, the total correction $Z_0+\delta Z$ remain approximately invariant within the well-sampled calibration domain. The effect of adopting this fixed-color baseline is examined in Section~\ref{sec:model_diagnostics}, where we compare it with models that retain the source-specific color dependence of L21.

Subtracting Equation~\eqref{eq:fixed_colour_baseline} from the observed parallax in Equation~\eqref{eq:raw_observation_model}, we define the baseline-corrected parallax
\begin{equation}
    r_i^{(k)}
    \equiv
    \hat{\varpi}_i^{(k)}-Z_{0,i}.
    \label{eq:baseline_corrected_parallax}
\end{equation}
It follows that
\begin{equation}
    r_i^{(k)}
    =
    \varpi_{c,k}
    + \delta Z
      \left(G_i,\nu_{{\rm eff},i};\theta\right)
    + \epsilon_i^{(k)}.
\end{equation}
We therefore adopt the likelihood
\begin{equation}
    r_i^{(k)}
    \sim
    \mathcal{N}\left[
        \varpi_{c,k}
        + \delta Z
          \left(G_i,\nu_{{\rm eff},i};\theta\right),
        \,
        V_{i,k}
    \right],
    \label{eq:likelihood}
\end{equation}
where
\begin{equation}
    V_{i,k} = \sigma_{\varpi,i}^{2} + s_k^{2}.
\end{equation}
Conditional on the model parameters, the member-star measurements are treated as independent. The term $s_k$ represents excess star-to-star scatter within cluster $k$, including contributions from underestimated formal uncertainties, crowding, unresolved binaries, imperfect membership assignments, and residual cluster depth. It does not represent a coherent astrometric offset shared by all members; such an offset is absorbed into $\varpi_{c,k}$. This conditional-independence approximation does not constitute a full model of the spatially correlated \textit{Gaia} errors, and uncertainties that neglect such correlations may therefore remain optimistic.

\subsection{Residual zero-point model}\label{sec:residaul_zp}
We assume that the color-fixed L21 baseline retains the dominant magnitude dependence of the parallax zero-point. The residual model is therefore intended to describe the color-dependent structure that remains relative to this baseline. Specifically, we include a color-only component and a magnitude--color interaction, but no independent magnitude-only term:
\begin{equation}
\begin{split}
\delta Z(G,\nu_{\rm eff}) =
&\sum_{q=1}^{Q} a_q C_q(\nu_{\rm eff}) \\
&+ \sum_{p=1}^{P}\sum_{q=1}^{Q}
b_{pq}
\left[M_p(G)C_q(\nu_{\rm eff})\right],
\end{split}
\label{eq:delta_z}
\end{equation}
where $C_q$ and $M_p$ are cubic B-spline contrasts in effective wavenumber and magnitude, respectively. Each marginal contrast is centered over the calibration sample, and every interaction contains both a centered magnitude contrast and a centered color contrast. Omitting a magnitude-only term is a deliberate constraint. It prevents the residual model from re-estimating the global magnitude dependence already encoded in L21. The magnitude--color interaction nevertheless allows the color dependence to vary with $G$, so the model is not required to adopt the same color correction at all magnitudes. In Section~\ref{sec:model_diagnostics}, we assess the effect of adding
magnitude--color interactions by comparing the fiducial model with a residual model containing only color terms.

Equation~(\ref{eq:delta_z}) can be written compactly as
\begin{equation}
    \delta Z(G,\nu_{\rm eff};\theta)
    = \sum_{j=1}^{M}\theta_j\Psi_j(G,\nu_{\rm eff}),
\end{equation}
where $M=Q+PQ$. 
We use cubic B-splines (degree $d=3$) in both dimensions. The magnitude basis has internal knots at $G=\{16,18\}$ mag. The color basis has four internal knots at $\nu_{\rm eff}=\{1.475,1.521,1.554,1.572\}\,\mu{\rm m}^{-1}$, corresponding to the 20th, 40th, 60th, and 80th percentiles of the calibration sample. After centering and removal of the redundant constant components, the model contains $P=5$ magnitude contrasts and $Q=7$ color contrasts. The pure-color component and the magnitude--color interactions therefore contain $M=Q+PQ=42$ coefficients in total.

\subsection{Prior distributions}
The cluster-specific parameter $\varpi_{{\rm c},k}$ is an effective parallax: after subtraction of the adopted global baseline, it represents the true mean parallax of cluster $k$ together with any residual component that is spatially coherent across its member stars. Although the cluster catalogs report mean parallaxes, these quantities are generally inferred from \textit{Gaia} astrometry and are therefore not independent of the measurements used in our likelihood.  Using these values as priors while fitting \textit{Gaia} parallaxes would double-count substantially overlapping astrometric information.

We therefore adopt the weakly informative prior for the effective parallax of each cluster,

\begin{equation}
\varpi_{{\rm c},k} \sim
\mathcal{N}(0,\tau_{\varpi}^2), \qquad
\tau_{\varpi}=10~{\rm mas}.
\end{equation}

The prior width is deliberately much larger than the parallax range of the
selected clusters and is therefore effectively flat over the region
supported by the data. Its purpose is to define a proper posterior and to
regularize the numerical optimization, rather than to provide distance
information. Because the residual model contains no free global constant and any cluster-common component is absorbed by $\varpi_{{\rm c},k}$, the model constrains the differential color--magnitude dependence relative to the adopted L21 baseline. It does not independently determine an additional absolute zero-point constant.

We allow for cluster-dependent excess star-to-star scatter through
\begin{equation}
s_k \sim {\rm HalfNormal}(\sigma_s), \qquad
\sigma_s=20~\mu{\rm as}.
\label{eq:scatter-prior}
\end{equation}
Here, $s_k$ enters the likelihood in quadrature with the formal parallax uncertainty. It accommodates residual error underestimation, imperfect membership selection, unresolved multiplicity, and other within-cluster effects not represented by the formal \textit{Gaia} uncertainties. The analysis of \citet{2021MNRAS.505.5978V} showed that formal DR3 parallax
uncertainties can be underestimated by approximately 10--20 per cent in dense cluster regions, and that spatially correlated systematics impose a floor of order $10~\mu{\rm as}$ on mean parallaxes of compact systems. We therefore adopt the more conservative scale $\sigma_s=20~\mu{\rm as}$. This term does not represent a spatially coherent zero-point offset, which is absorbed by $\varpi_{{\rm c},k}$.

Finally, we assign independent zero-centered Gaussian priors to the residual
spline coefficients,
\begin{equation}
\theta_j \sim \mathcal{N}(0,\sigma_\theta^2), \qquad
\sigma_\theta=20~\mu{\rm as}.
\label{eq:coefficient-prior}
\end{equation}
This prior regularizes the amplitude of the flexible residual correction and suppresses poorly constrained excursions in sparsely populated regions of the color--magnitude plane. Its scale is motivated by the residual parallax structures of order $10$--$30~\mu{\rm as}$ that remain after
application of the L21 correction in D26 study. 

\subsection{Posterior inference}

Up to an additive constant, the log-posterior is
\begin{equation}
\begin{split}
\ln p(\boldsymbol{\phi}\mid\mathcal{D}) =
&-\frac{1}{2}\sum_{k=1}^{K}\sum_{i=1}^{N_k}
\left[
\ln V_{i,k} +
\frac{
\left[r_i^{(k)}-\varpi_{c,k}-\delta Z_i(\theta)\right]^2
}{V_{i,k}}
\right] \\
&-\frac{1}{2}\sum_{k=1}^{K}
\frac{\varpi_{c,k}^{2}}{\tau_\varpi^2}
-\frac{1}{2}\sum_{j=1}^{M}
\frac{\theta_j^2}{\sigma_\theta^2}
-\frac{1}{2}\sum_{k=1}^{K}
\frac{s_k^2}{\sigma_s^2},
\end{split}
\label{eq:logposterior}
\end{equation}
where terms that are constant with respect to the parameters have been omitted. We adopt the maximum a posteriori (MAP; \citealt{1994ITSAP...2..291G}) solution as the primary estimator. Full posterior sampling with conventional Markov Chain Monte Carlo methods (MCMC; \citealt{1953JChPh..21.1087M,1970Bimka..57...97H,2017ARA&A..55..213S}) would be computationally
expensive for the large number of cluster-specific parameters and B-spline coefficients considered here. We therefore minimize the negative log-posterior with the L-BFGS-B algorithm \citep{Zhu1997Algorithm7L}, supplying analytic gradients and enforcing $s_k>0$ through a logarithmic parameterization. 

\begin{figure}
    \centering
    \subfigure{\includegraphics[width=1.0\linewidth]{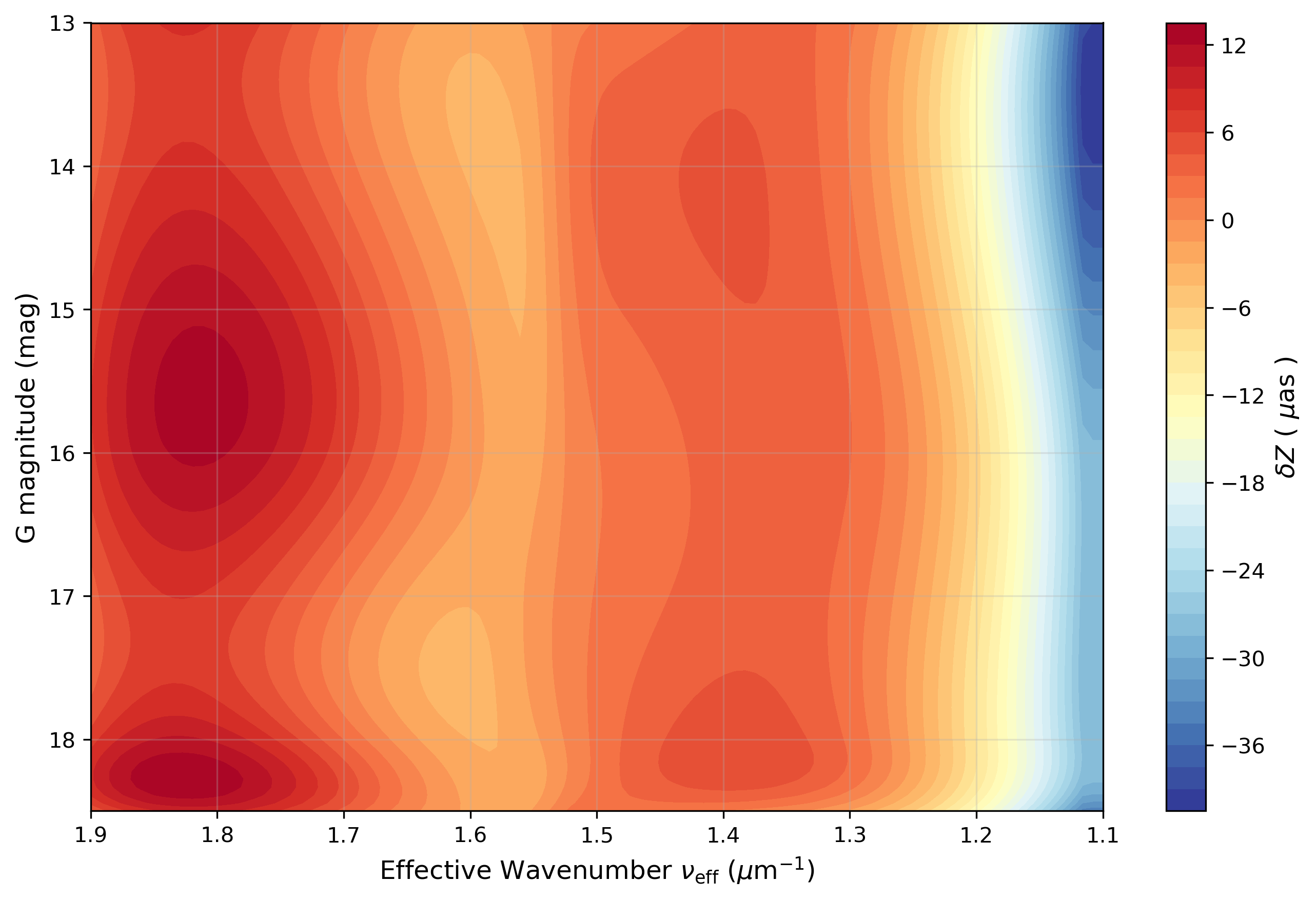}}
    \caption{MAP residual parallax zero-point correction $\delta Z(G,\nu_{\rm eff})$ inferred from the globular-cluster sample. The correction is modestly positive over most of the intermediate and blue color domain, whereas it decreases sharply towards the reddest sources ($\nu_{\rm eff}\lesssim1.2~\mu{\rm m}^{-1}$).}
    \label{fig:delta_z_surface}
\end{figure}

\begin{figure*}
    \centering
    \subfigure{\includegraphics[width=0.32\linewidth]{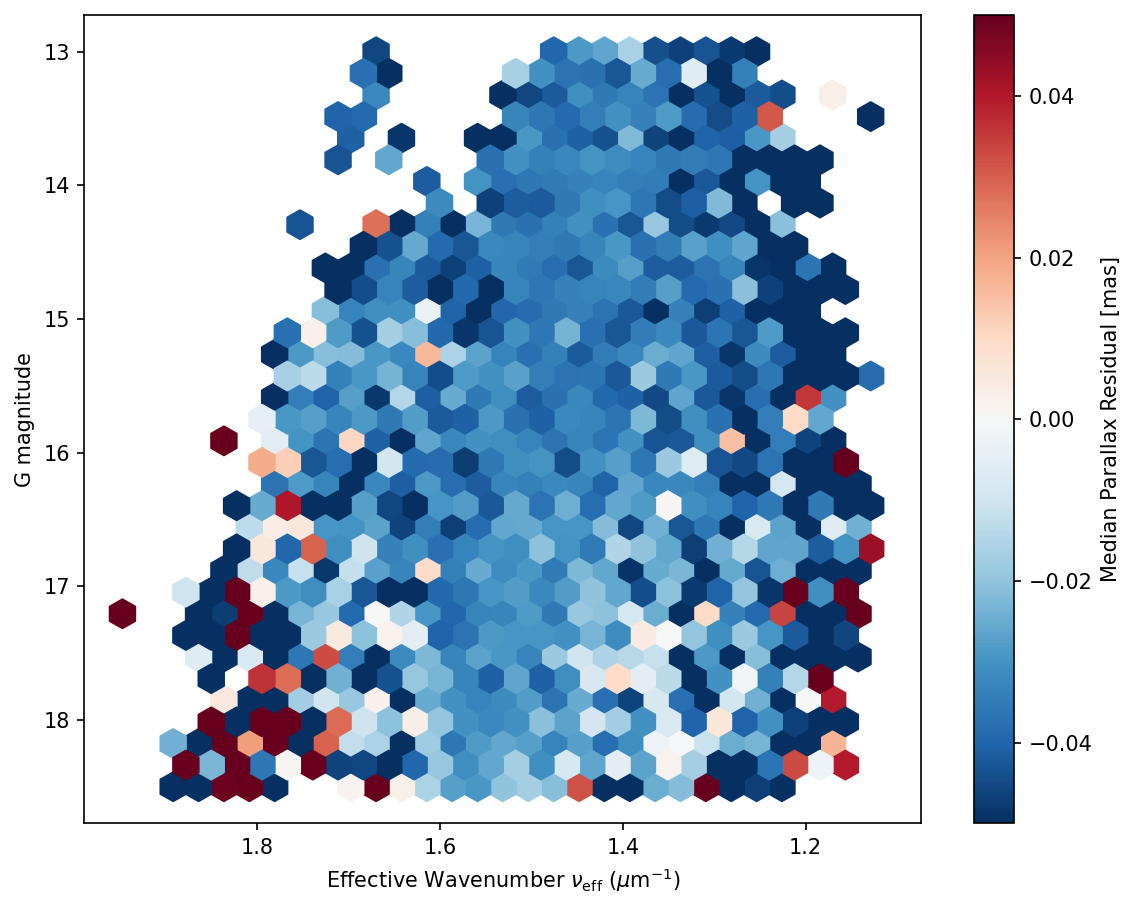}}
    \subfigure{\includegraphics[width=0.32\linewidth]{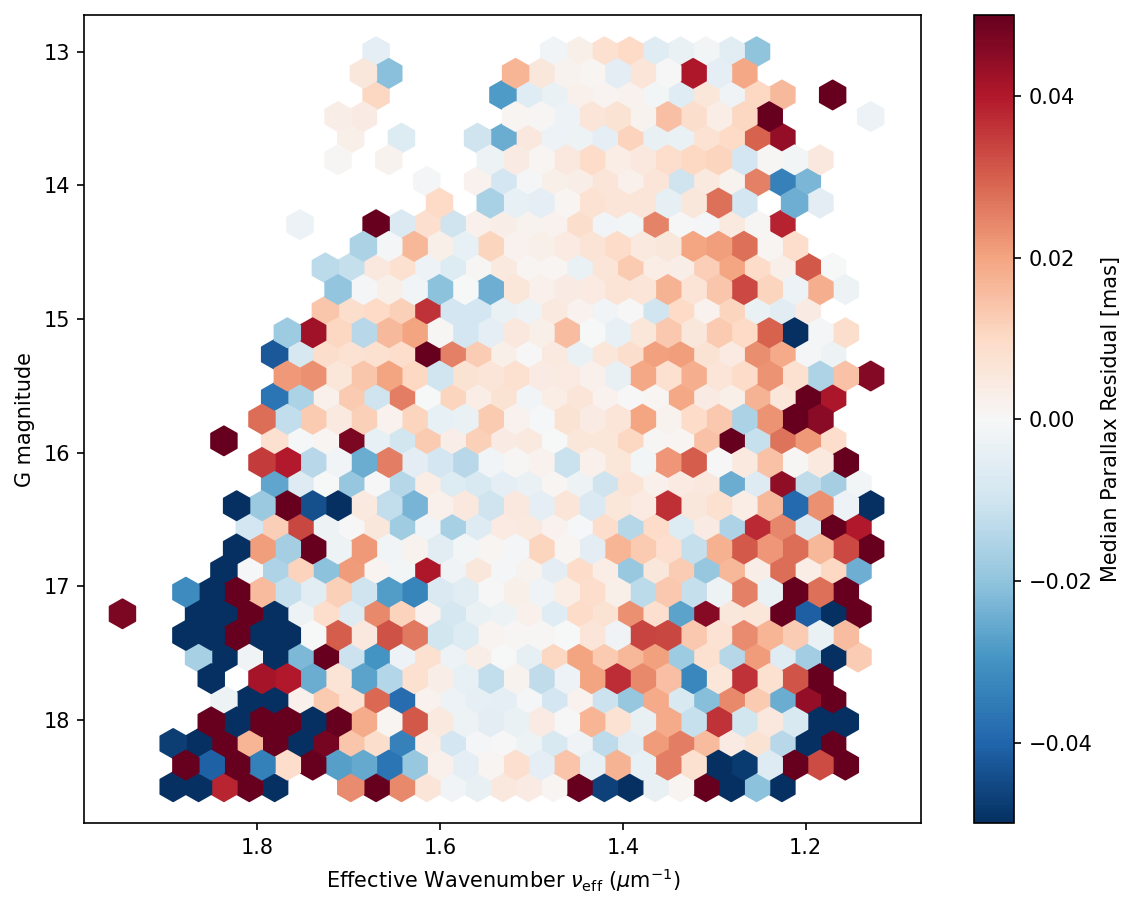}}
    \subfigure{\includegraphics[width=0.32\linewidth]{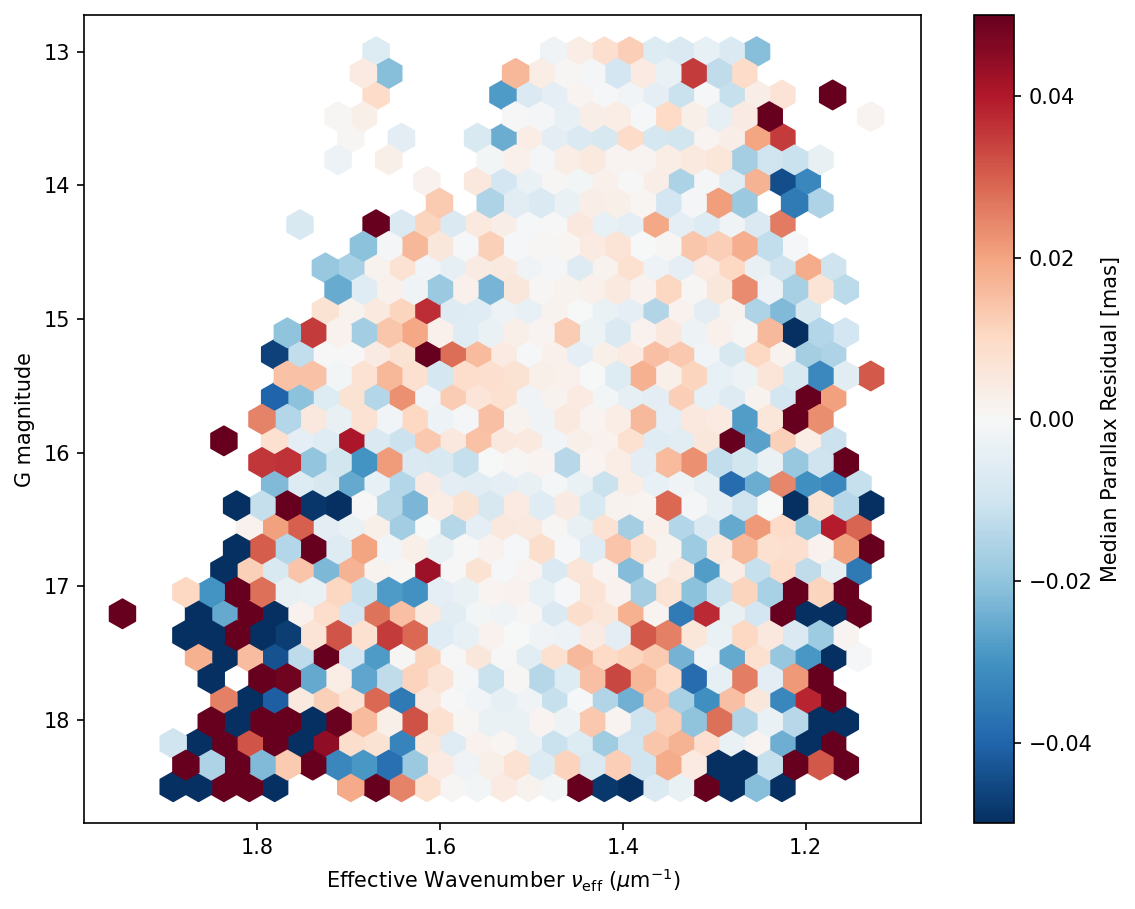}}
    \caption{Parallax residuals of GC members in the $(G,\nu_{\rm eff})$ plane. The panels show the DR3 residuals (left), L21-corrected residuals (middle), and those after the fixed-color L21 plus $\delta Z(G,\nu_{\rm eff})$ (right), respectively. The same model-inferred cluster parallax is used in all three panels. Each colored hexagonal cell shows the median member-star residual in that cell.}
    \label{fig:cmd_residual_maps}
\end{figure*}

\begin{figure}
    \centering
    \subfigure{\includegraphics[width=1.0\linewidth]{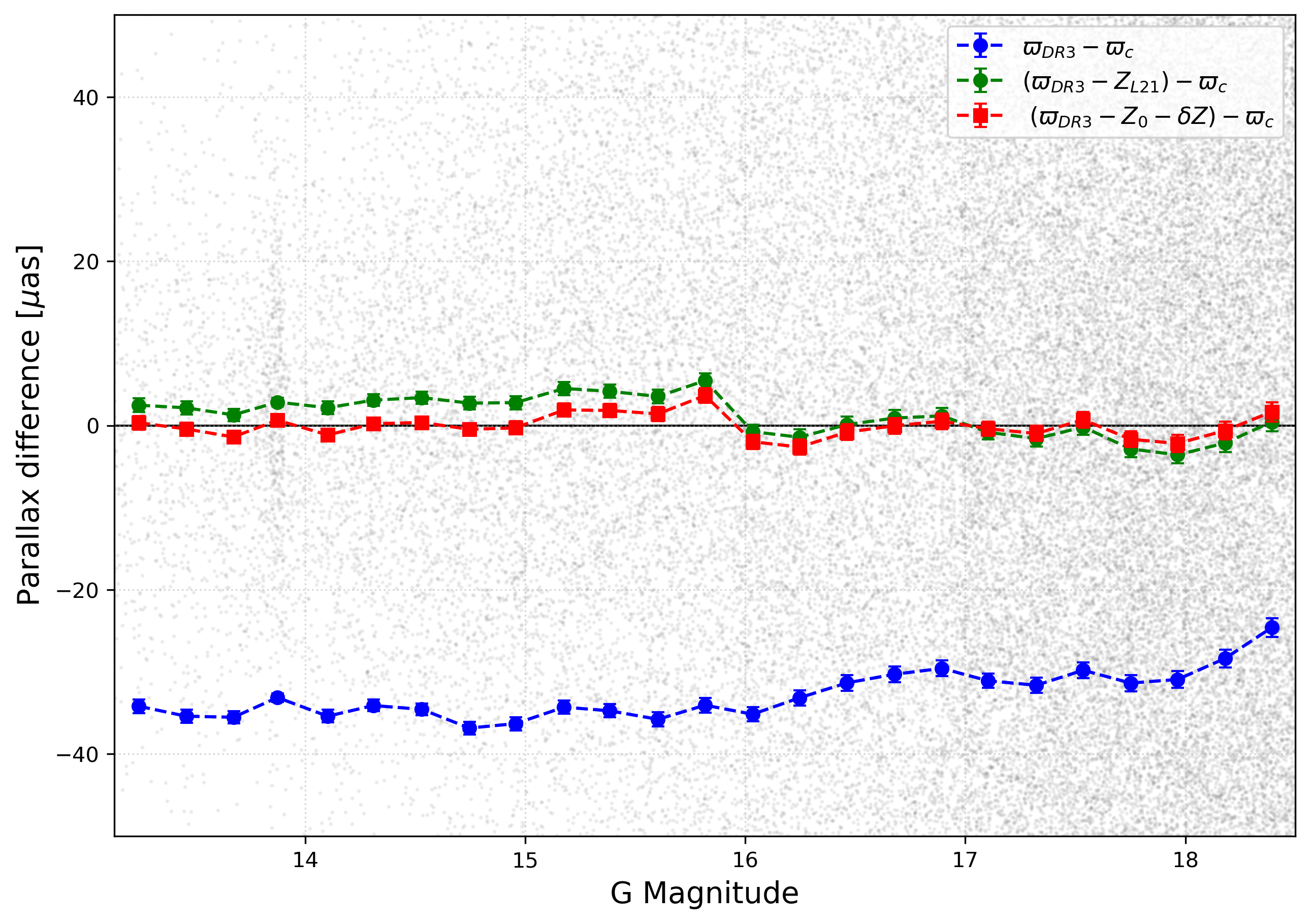}}
    \subfigure{\includegraphics[width=1.0\linewidth]{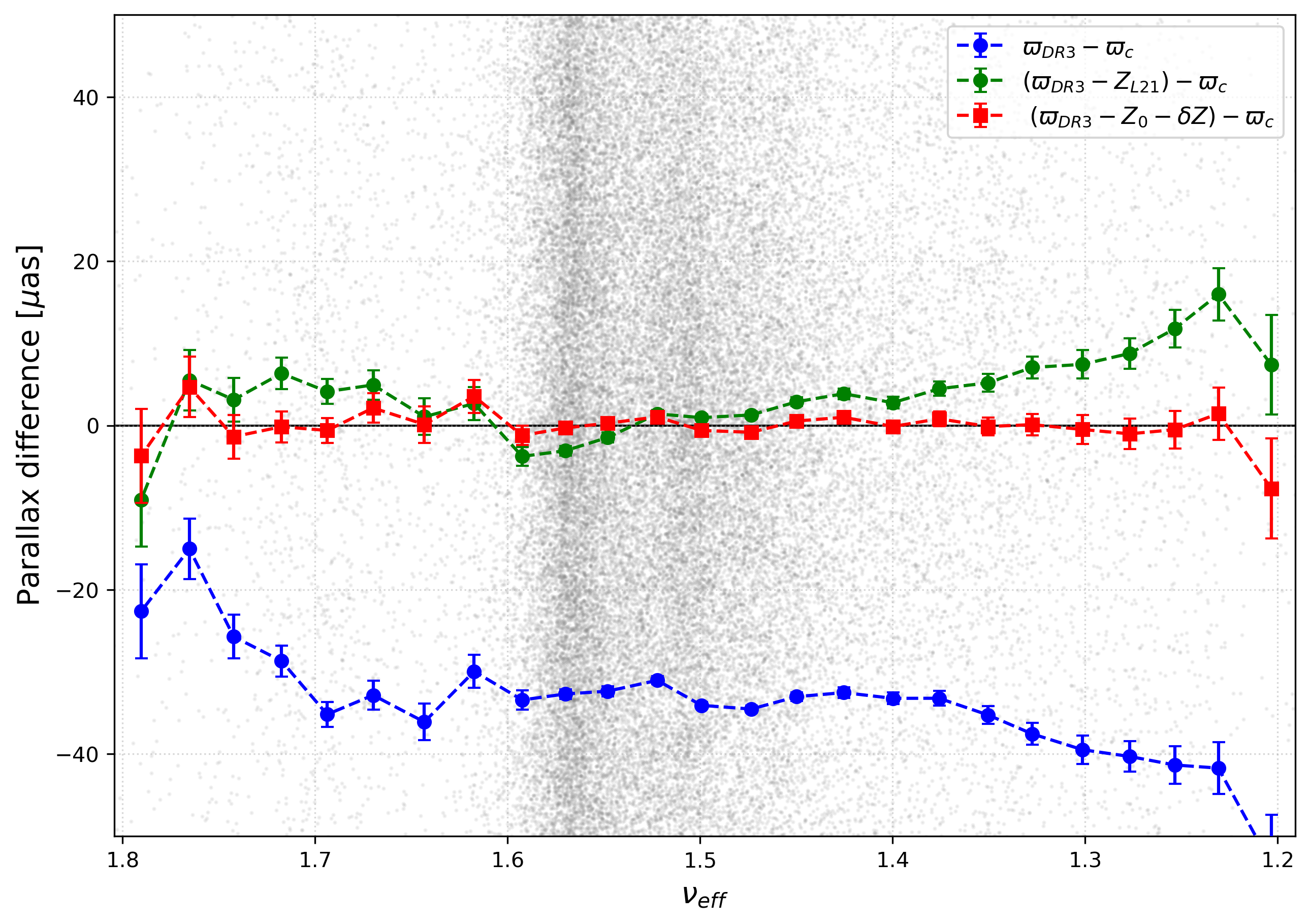}}
    \caption{Binned parallax residuals before and after applying the fitted residual correction, plotted against $G$ magnitude (top), and effective wavenumber $\nu_{\rm eff}$ (bottom). Blue, green, and red curves denote the raw \textit{Gaia} DR3 residuals, the residuals after the adopted L21 baseline, and those after the L21 baseline plus $\delta Z(G,\nu_{\rm eff})$, respectively. The final curve is evaluated relative to the fitted effective cluster parallax.}
    \label{fig:plxdiff_trend}
\end{figure}

\section{Results} \label{sec:result}

We fitted the hierarchical residual zero-point model to 107,222 member stars in 138 GCs. The model jointly determines the coefficients of $\delta Z(G,\nu_{\rm eff})$, the effective cluster parallaxes $\{\varpi_{c,k}\}$, and the cluster-specific extra-scatter terms $\{s_k\}$. All fitted quantities are MAP point estimates. 

Figure~\ref{fig:delta_z_surface} summarizes the residual parallax zero-point correction inferred from the GC sample. The MAP solution identifies a predominantly color-dependent structure relative to the adopted L21 baseline: $\delta Z$ is modest over most of the sampled
color--magnitude domain, but decreases sharply towards the reddest sources, reaching approximately $-50~\mu{\rm as}$ at $\nu_{\rm eff}\lesssim1.15~\mu{\rm m}^{-1}$. The magnitude dependence provides a smooth modulation of this broad color trend. The corresponding spline coefficients remain at the tens-of-microarcsecond level, consistent with the adopted regularization.




Figure~\ref{fig:cmd_residual_maps} compares the parallax residuals across the $(G,\nu_{\rm eff})$ plane before correction, after the L21 correction, and after our fixed-color L21 baseline plus
$\delta Z(G,\nu_{\rm eff})$, respectively. In all three panels, the same model-inferred cluster parallax, $\varpi_{c,k}$, is subtracted from each member star. The raw residuals show a predominantly negative residual across the populated region of the diagram. The full L21 correction removes much of this common offset, but color--magnitude-dependent structure remains. With the proposed combined correction, the median residuals are more uniformly distributed around zero over the densely sampled region, although localized positive and negative residuals persist, particularly towards the sparsely populated boundaries. These maps provide a two-dimensional view of the differential trends examined below.

Figure~\ref{fig:plxdiff_trend} compares the binned parallax residuals relative to the same fitted effective cluster parallaxes. In the top panel, the uncorrected \textit{Gaia} DR3 parallaxes show a predominantly negative offset of about $25$--$40~\mu{\rm as}$. The adopted color-fixed
L21 baseline removes most of this offset. Adding $\delta Z(G,\nu_{\rm eff})$ reduces the smaller variations at brighter magnitudes. The faintest bin ($G >$ 18.5) is particularly uncertain and is not used to assess the correction.
The bottom panel of Fig.~\ref{fig:plxdiff_trend} shows a clearer improvement with color. After the L21 baseline, the residuals rise towards the red end, reaching approximately $10$--$20~\mu{\rm as}$ at $\nu_{\rm eff}\simeq1.25$--$1.35~\mu{\rm m}^{-1}$. The fitted correction largely removes this trend, leaving binned residuals close to zero across most of the sampled color range. Thus, the main gain is the suppression of the residual color dependence.

\section{Validation}\label{sec:validation}

We assess the stability of the residual correction using five-fold cross-validation \citep{rodriguez2009sensitivity}. Complete GCs, rather than individual members, are assigned to folds. In each iteration, The spline basis and coefficients are determined from the other four folds, and the fitted $\delta Z$ is evaluated for the held-out clusters. Each fold contains 27--28 test clusters; thus, all 138 GCs and their 107,222 retained members are evaluated once by a model fitted without the corresponding cluster.

For all three correction prescriptions, we use the same fixed cluster parallax, taken from the model fitted to the full GC sample. Member residuals are then calculated relative to that cluster estimate. Residuals are averaged first within each cluster and magnitude or color bin, and then across the contributing clusters, giving each represented cluster equal weight.

\begin{figure}
    \centering
    \includegraphics[width=\columnwidth]{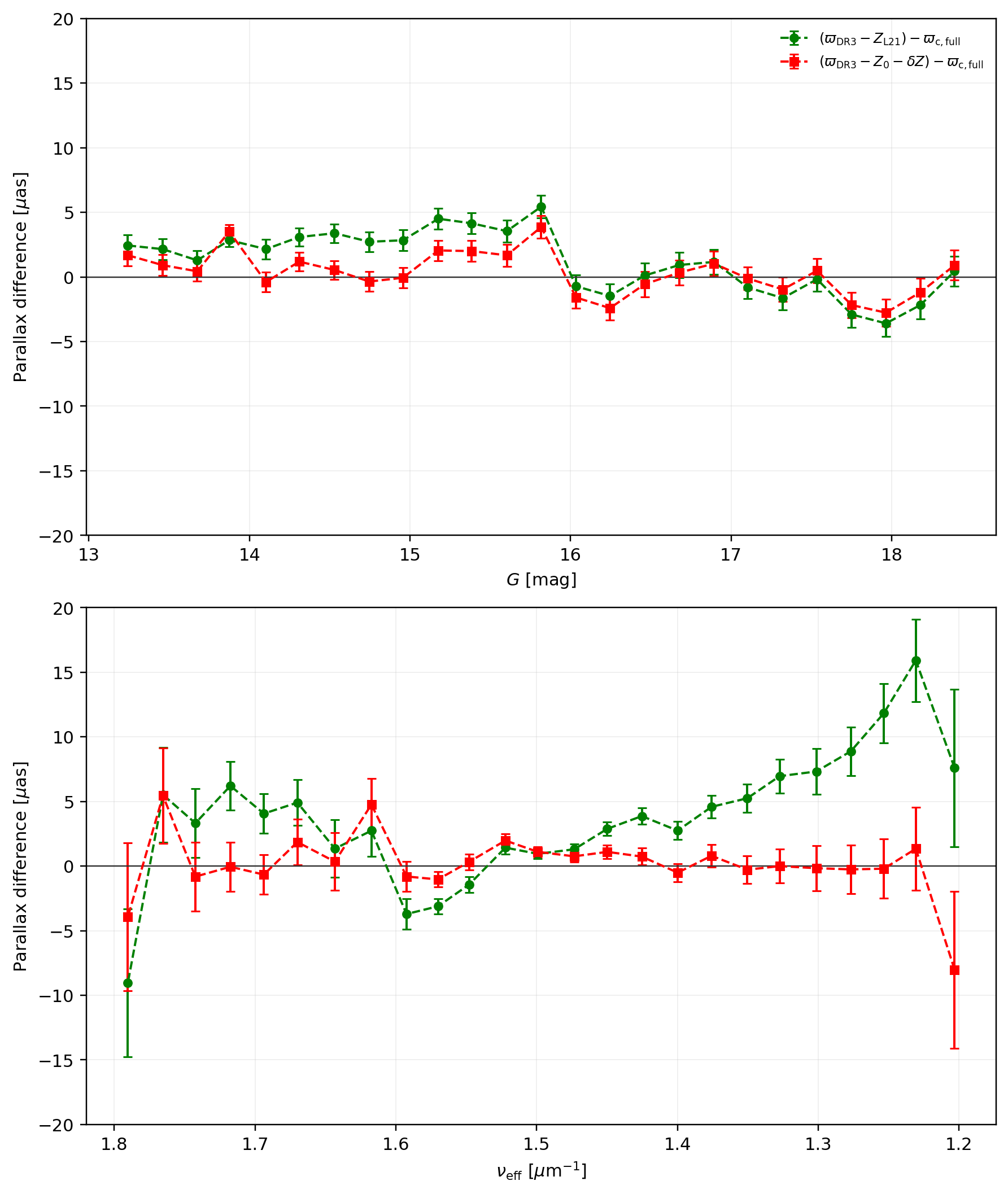}
    \caption{Out-of-fold, cluster-centered parallax residuals as functions of $G$ (top) and $\nu_{\rm eff}$ (bottom) in five-fold cluster-grouped validation. Green circles show the full L21 correction, and red squares the fixed-color L21 baseline plus $\delta Z$. For each held-out cluster and each prescription, an effective parallax and additional scatter are estimated from all retained members before calculating their residuals. Points are means of the inverse-variance-weighted cluster-bin residuals, with equal weight assigned to each contributing cluster. Error bars show the
    standard error across clusters.}
    \label{fig:gc_validation}
\end{figure}

Figure~\ref{fig:gc_validation} compares the out-of-fold differential residuals for the full L21 correction (green), and the fixed-color L21 baseline combined with $\delta Z$ (red). Relative to full L21, the combined prescription reduces the RMS of the magnitude-bin means from $2.65$  to $1.67\,\mu{\rm as}$ and that of the color-bin means from $6.17$ to $2.44\,\mu{\rm as}$. The improvement on clusters is more effective in color than in magnitude, especially at $\nu_{\rm eff}\simeq1.2$--$1.5~\mu{\rm m}^{-1}$.

Estimating a separate parallax for each cluster removes its common offset and can leave the raw residual trend relatively flat even when the absolute parallaxes are biased. This experiment therefore tests whether the fitted color--magnitude dependence transfers to unseen GCs; it does not independently validate the absolute parallax zero-point or predict individual-member residuals using an independently known cluster parallax.

\section{Discussion} \label{sec:discussion}

\subsection{The residual-model formalism}
\label{sec:model_diagnostics}
We tested four model specifications obtained by combining either the full color-dependent or the fixed-color L21 baseline with either a color-only or a magnitude--color interactions. Within the well-sampled GC color--magnitude domain, all four specifications yield nearly identical residual statistics. Their standardized residual RMS values differ by less than $3\times10^{-4}$, and the differences among the one- and two-dimensional residual-trend metrics are typically below $1~\mu{\rm as}$. Cluster-grouped cross-validation likewise provides no decisive preference for a single specification.

The inferred correction is therefore insensitive to the adopted parameterization within the region supported by the calibration data. The model surfaces diverge mainly near sparsely populated boundaries, where the correction is weakly constrained and should not be extrapolated. We retain the fixed-color L21 baseline with magnitude--color interactions as the fiducial model because it allows the residual color dependence to vary with magnitude. Its marginally smaller residual metrics, however, do not establish a statistically unique solution. Detailed comparisons are presented in Appendix~\ref{app:ablation}.

\subsection{Investigation of the open clusters}
Open clusters (OCs) are examined only as an exploratory comparison because their heterogeneous member catalogs exhibit substantially larger scatter. We explored several published OCs member catalogs \citep{2014A&A...564A..79D,2020A&A...640A...1C,2023ApJS..267...34H,2024A&A...686A..42H,2025AJ....170..288L}. We do not treat them as an independent validation set. The catalogs differ in membership construction, cross-matching strategy, cluster compactness, and contamination level, and some exhibit within-cluster parallax trends far larger than the GC-derived correction. Moreover, their cluster means are themselves estimated from the same \textit{Gaia} parallaxes, so they provide no external absolute reference. The response to the GC-derived correction is consequently catalog dependent: some color trends decrease, while the magnitude trends generally do not improve. Representative example is retained in Appendix~\ref{app:oc_catalogs} as a limitation and domain-transfer diagnostic rather than as evidence used to select the model.

\subsection{Application and limitations}

For a \textit{Gaia} DR3 source with a five-parameter astrometric solution, we evaluate the L21 zero-point at the same fixed wavenumber, $\nu_0=1.59~\mu{\rm m}^{-1}$, while retaining its dependence
on $G$ and ecliptic latitude. The source's measured $\nu_{\rm eff}$ enters only the fitted residual term, $\delta Z_{\rm GC}$. Thus, the correction \footnote{The fitted model coefficients, Python implementation of Equation~\eqref{eq:application}, and an editable application example are publicly available at \url{https://github.com/yeding/gaia-dr3-gc-zpt}.} applied in this work is
\begin{equation}\label{eq:application}
\varpi_{\rm corr} = \varpi_{\rm DR3} - Z_{\rm L21}(G,\nu_0,\beta) - \delta Z_{\rm GC}(G,\nu_{\rm eff})
\end{equation}
Because $\delta Z_{\rm GC}$ was fitted relative to this fixed-color L21 baseline, it should not be added to L21 evaluated at each source's measured $\nu_{\rm eff}$. 

The empirical validity of Equation~\eqref{eq:application} is restricted to five-parameter sources occupying the color--magnitude domain sampled by the GC calibration members (13 $<G<$ 18.5, 1.1$<\nu_{\rm eff}<$1.9). Predictions close to the sample boundaries should likewise be treated with caution. The model is not applicable to six-parameter astrometric solutions, for which the Gaia calibration depends on a pseudo-color and follows a different zero-point behavior.
The model was calibrated using high-probability members of Galactic GCs and is intended primarily for differential analyses of comparable cluster-member samples. 

Finally, the cluster parallaxes used in the calibration were inferred from the same \textit{Gaia} members, and the fitted surface represents a MAP estimate rather than a fully sampled posterior predictive correction. The grouped diagnostic therefore supports an improvement in within-cluster differential consistency, but does not provide an independent validation of the absolute parallax or distance scale. We consequently recommend the model as a differential refinement for GC-like samples within the well-supported calibration domain. Absolute cluster distances, applications near the domain boundaries, and use with other stellar populations require independent validation and additional systematic-error allowances.

\section{Summary} \label{sec:summary}

\textit{Gaia} DR3 parallax zero point exhibits a complex joint dependence on magnitude, color, and sky position. A smooth global prescription cannot necessarily capture all of this structure, particularly in crowded regions and in compact systems such as Galatic clusters, where crowding, source-property distributions, and small-scale calibration variations differ from those of the reference samples used to construct the global model. Astrometric errors among members of the same cluster are also spatially correlated. Consequently, uncertainties derived by treating individual member measurements as independent---including formal uncertainties on cluster-mean parallaxes---are likely to be overly optimistic. 

In this work, we used 107,222 five-parameter \textit{Gaia} DR3 sources in 138 Galactic GCs to characterize residual variations in the parallax zero-point with magnitude and effective wavenumber. Our hierarchical model assigns each cluster an effective parallax and an additional scatter term, while describing the shared residual correction with centered cubic B-spline color terms and magnitude--color interactions. The L21 baseline is evaluated at a fixed reference wavenumber, allowing the residual model to represent the remaining source-dependent color structure explicitly.

The fitted correction substantially suppresses the color-dependent residuals remaining after the adopted L21 baseline, particularly toward the red end of the GC sample. The mean residuals are close to zero across most of the well-sampled color--magnitude domain, although some structure persists near the faint and sparsely populated boundaries. Because the effective cluster parallaxes and the correction are inferred from the same member data, these results demonstrate an
improvement in internal differential consistency rather than an independent calibration of the absolute parallax zero-point.

We also performed five cluster-grouped fits in which complete GCs were excluded from the estimation of the spline coefficients. Relative to the full L21 prescription, the fixed-color L21 baseline combined with $\delta Z$ reduces the RMS of the magnitude-bin residual means from
$2.65$ to $1.67\,\mu{\rm as}$ and that of the color-bin means from $6.17$ to $2.44\,\mu{\rm as}$. The reduction is more pronounced in color than in magnitude, particularly at $\nu_{\rm eff}\simeq1.2$--$1.5~\mu{\rm m}^{-1}$. Because the cluster parallaxes used in this diagnostic are taken from the full-sample fit, the experiment tests the stability of the residual spline correction, but does not constitute a fully independent validation of the complete model or its absolute zero-point scale.

Exploratory applications to heterogeneous OC catalogs likewise do not show uniform transferability and are therefore not used as formal validation. The GC-derived correction should consequently be regarded as a differential refinement for five-parameter DR3 sources within the GC color--magnitude domain, and not as a general replacement for the L21 correction. Independent distance anchors and more homogeneous external samples are required to
establish its absolute calibration and its applicability to other stellar populations.

Future \textit{Gaia} DR4 astrometry is expected to benefit from a longer observational baseline and revised calibrations, and we hope that the complex color-, magnitude-, and position-dependent systematics evident in DR3 will be substantially reduced. Nevertheless, residual zero-point structure will need to be assessed empirically, especially for dense stellar systems. The framework developed here is not tied to a particular DR3 correction surface: a DR4 baseline can be substituted, and the residual color--magnitude dependence can be re-estimated from high-probability GC members while accounting for cluster-specific parallaxes and excess scatter. The present analysis therefore provides both a DR3 differential correction within its calibrated domain and a general strategy for evaluating and refining the parallax zero-point of GCs in \textit{Gaia} DR4.

\begin{acknowledgments}
This work was supported by the National Natural Science Foundation of China (Grant No. 12673081, 12603104), the National Key R$\&$D Program of China (Grant No. 2025YFA1614104),  the Strategic Priority Research Program of the Chinese Academy of Sciences (Grant No. XDA0350205), the Youth Innovation Promotion Association of the Chinese Academy of Sciences (Certificate No. 2022259), the Talent Program of the Shanghai Branch of the Chinese Academy of Sciences (Grant No. CASSHB-QNPD-2023-016), and the International Partnership Program of the Chinese Academy of Sciences (Grant No. 018GJHZ2025032FN). We also acknowledge support from the China Manned Space Project (Grant Nos. CMS-CSST-2021-A12 and CMS-CSST-2021-B10). This work has made use of data from the European Space Agency (ESA) mission \textit{Gaia} (\href{https://www.cosmos.esa.int/gaia}{https://www.cosmos.esa.int/gaia}), processed by the \textit{Gaia} Data Processing and Analysis Consortium (DPAC, \href{https://www.cosmos.esa.int/web/gaia/dpac/consortium}{https://www.cosmos.esa.int/web/gaia/dpac/consortium}). Funding for the DPAC has been provided by national institutions, in particular the institutions participating in the \textit{Gaia} Multilateral Agreement. This research has made use of the VizieR catalogue access tool and the cross-match service provided by CDS, Strasbourg.

\end{acknowledgments}

\appendix

\section{Sensitivity to the residual-model specification}
\label{app:ablation}

We compare four specifications formed by combining the full or fixed-color L21 baseline with a color-only or magnitude--color residual model. Because these models partition the L21 baseline and the residual correction differently, we compare their total corrections and cluster-balanced residuals rather than individual spline coefficients.

\begin{figure*}
    \centering
    \includegraphics[width=\linewidth]{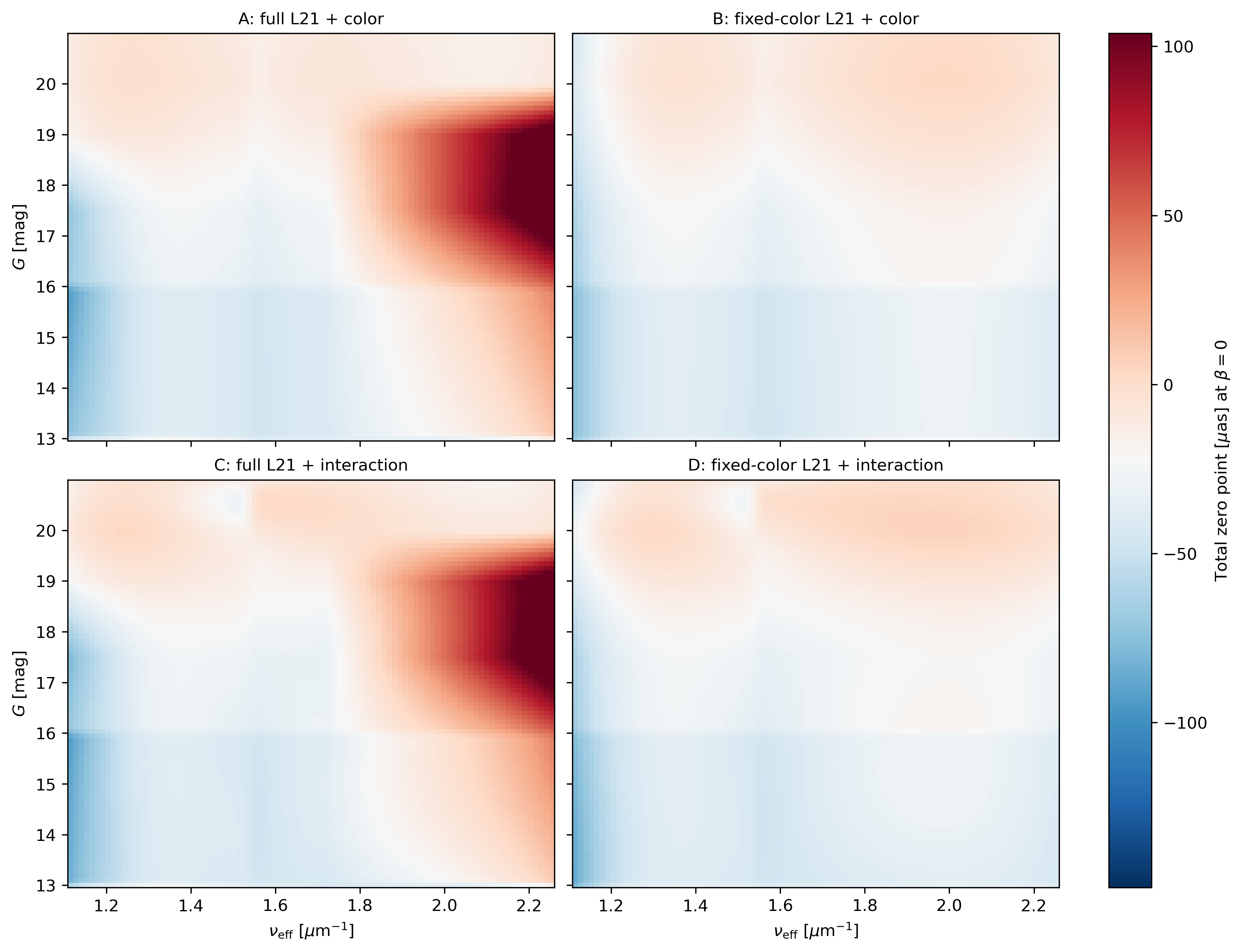}
    \caption{Total parallax zero-point corrections predicted by the four model specifications at $\beta=0$. All panels use the same color scale. The total surfaces are more directly comparable than the raw B-spline coefficients because the definition of the residual component depends on the adopted L21 baseline.}
    \label{fig:ablation_surfaces}
\end{figure*}

Figure~\ref{fig:ablation_surfaces} compares the total zero-point corrections predicted by four model specifications. Models A and C use the full source-dependent L21 correction as their baseline, whereas Models B and D evaluate L21 at the fixed reference wavenumber $\nu_{\rm ref}=1.59~\mu{\rm m}^{-1}$. Models A and B restrict the additional residual correction to a color-only component. Models C and D additionally include magnitude--color interactions, allowing the
residual color dependence to vary with $G$. Model D therefore corresponds to our fiducial specification. Within the densely populated region of the GC color--magnitude distribution, all four models recover broadly similar correction patterns. Models A and C, which retain the source-specific color dependence of L21, produce nearly identical surfaces over most of the domain; their magnitude--color interactions mainly affect regions near the boundaries. Models B and D likewise remain similar across the well-supported region, indicating that the additional interaction terms
introduce only moderate changes where the data provide strong constraints. 

\begin{figure}
    \centering
    \includegraphics[width=\linewidth]{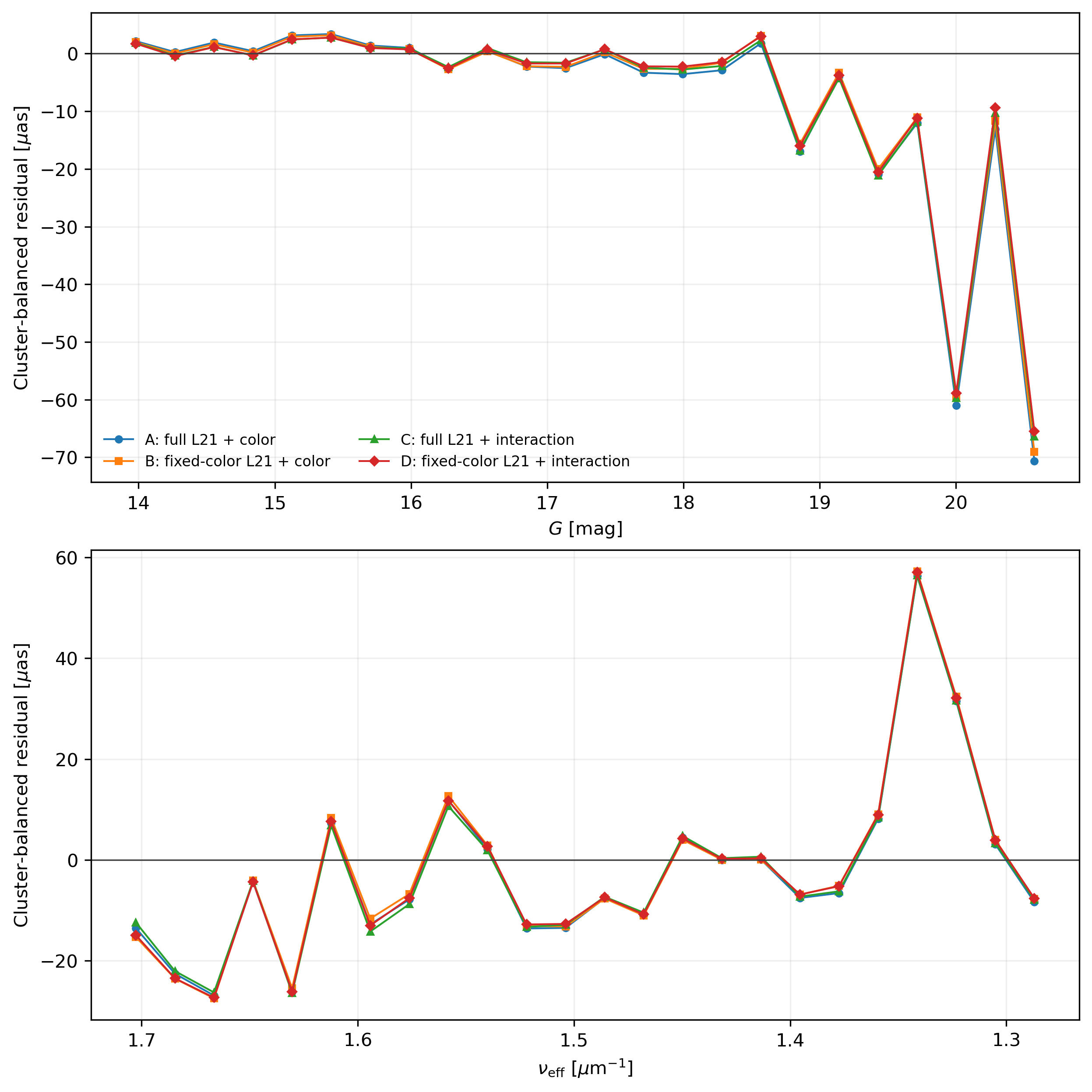}
    \caption{Cluster-balanced residual trends as functions of magnitude and effective wavenumber for the four model specifications. Each represented cluster receives equal weight within a bin.}
    \label{fig:ablation_trends}
\end{figure}

Figure~\ref{fig:ablation_trends} further shows that the corresponding one-dimensional residual trends are nearly indistinguishable within the calibrated domain. The present data therefore do not provide strong statistical evidence for one of the four specifications based on residual statistics alone. We nevertheless adopt Model D---the fixed-color L21 baseline combined with color and magnitude--color residual terms---as the fiducial model. This formulation assigns the source-dependent color behavior explicitly to $\delta Z$ and allows that behavior to change with magnitude, while avoiding the simultaneous use of the full L21 color dependence and an additional fitted color correction. The divergence among the models near the sample boundaries also motivates restricting applications of the fiducial correction to the well-supported GC color--magnitude domain.


\section{Open-cluster catalog diagnostics}
\label{app:oc_catalogs}

As an exploratory test of the applicability of the GC-derived correction to a different cluster population, we apply it to the open-cluster members compiled by \citeauthor{2020A&A...640A...1C} (\citeyear{2020A&A...640A...1C}, hereafter CG20). The spline coefficients inferred from the GC sample are held fixed throughout this exercise. For each OC and each correction prescription, a separate mean parallax is estimated from its retained members and removed before the residual trends are calculated. The resulting comparison therefore tests only the within-cluster dependence on magnitude and color; it
contains no independent information about either the cluster distances or the absolute parallax zero-point.

\begin{figure}
    \centering
    \includegraphics[width=\linewidth]{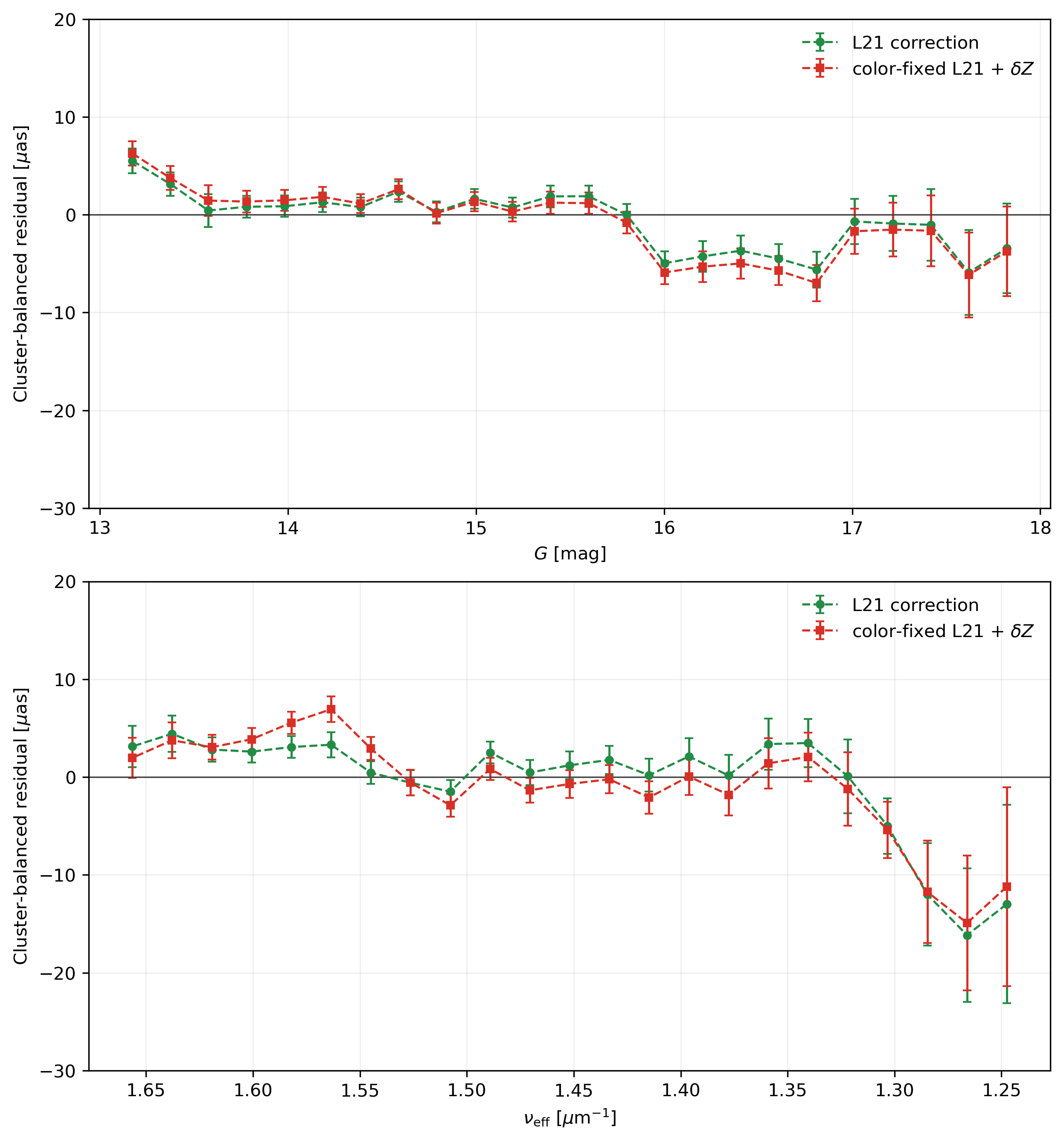}
    \caption{ Exploratory application of the GC-derived correction to the CG20 sample. The top and bottom panels show cluster-centered parallax residuals as functions of $G$ and
    $\nu_{\rm eff}$, respectively. Green circles represent the full source-dependent L21 correction, while red squares represent the fixed-color L21 baseline combined with the GC-derived
    $\delta Z(G,\nu_{\rm eff})$. For each prescription, a separate mean parallax is estimated and removed for every OC. The points are cluster-balanced bin means, and the error bars show the
    standard errors across the contributing clusters. This comparison probes only differential trends within the CG20 sample and is not used as an independent validation of the absolute parallax
    zero-point.}
    \label{fig:cantat20_trend}
\end{figure}

Figure~\ref{fig:cantat20_trend} compares the full, source-dependent L21 correction with the fixed-color L21 baseline combined with the GC-derived $\delta Z$. The two prescriptions produce broadly similar
magnitude-dependent residuals. Both remain close to zero at $G\lesssim15.8$, followed by a negative feature around $G\simeq16$--17~mag. In several of these bins, the addition of $\delta Z$ produces a slightly more negative residual rather than a systematic improvement. The color-dependent trends likewise show a mixed response. The GC-derived correction changes the residuals by only a few
$\mu{\rm as}$ over most of the sampled range and does not uniformly reduce their amplitude. Both prescriptions show increasingly negative residuals toward the reddest boundary, where the uncertainties also become substantially larger. The CG20 application therefore provides no clear evidence that the GC-derived correction transfers unchanged to OCs. We consequently retain it only as an illustrative diagnostic and do not use the OC sample for formal model validation.

\bibliography{myref}{}
\bibliographystyle{aasjournalv7}



\end{document}